\documentclass[showpacs,amsmath,amssymb,showkeys,pre]{revtex4}
\usepackage{graphicx}

\newcommand{\beq}{\begin{equation}}
\newcommand{\eeq}{\end{equation}}
\newcommand{\beqa}{\begin{eqnarray}}
\newcommand{\eeqa}{\end{eqnarray}}
\newcommand{\non}{\nonumber}

\begin{document}

\title{On the formal equivalence of the TAP and thermodynamic methods \\
in the SK model}

\author{Andrea Cavagna, Irene Giardina, Giorgio Parisi}

\affiliation{Center for Statistical Mechanics and Complexity, INFM Roma ``La Sapienza'' and \\
Dipartimento di Fisica, Universit\`a di Roma ``La Sapienza'', Piazzale Aldo Moro 2, 00185 Roma, 
Italy}

\author{Marc M\'ezard}
\affiliation{Laboratoire de Physique Th\'eorique et Mod\`eles Statistiques \\
Universit\'e\ Paris Sud, Bat. 100, 91405 Orsay Cedex, France}

\date{October 30, 2002}

\begin{abstract}
We revisit two classic Thouless-Anderson-Palmer (TAP) studies of the
Sherrington-Kirkpatrick model [Bray A J and Moore M A 1980 {\it J. Phys. C}
{\bf 13} L469; De Dominicis C and Young A P, 1983 {\it J. Phys. A} {\bf 16}
2063]. By using the Becchi-Rouet-Stora-Tyutin (BRST) supersymmetry, we prove the
general equivalence of TAP and replica partition functions, and show that the
annealed calculation of the TAP complexity is formally identical to the
quenched thermodynamic calculation of the free energy at one step level 
of replica symmetry breaking. The complexity we obtain by means of
the BRST symmetry turns out to be considerably smaller than the previous 
non-symmetric value. 
\end{abstract}


\maketitle

\section{Introduction}

The static properties of mean-field spin-glasses have been investigated in the
past mainly by means of two different approaches: standard thermodynamics,
which through the replica method \cite{ea,sk,sk2,1rsb,2rsb} aims to compute
the  partition function and the
equilibrium free energy density of the system, and the
Thouless-Anderson-Palmer (TAP) method \cite{tap,bm,bm1,bm2,dd1,innocent2,ddy},
which introduces a mean-field free energy $F_{TAP}$, function of the set $m$
of local magnetizations, $m\equiv\{m_i\}_{i=1}^N$. The local minima of
$F_{TAP}(m)$ are identified with the metastable states of the system, which at
the mean-field level are well defined. At low temperatures the total number
$\cal N$ of minima of the TAP free energy becomes exponentially large
\cite{bm,bm1,dd1}, and the density of states is given by, \beq \rho(f) \sim
e^{N\Sigma(f)} \ .
\label{density}
\eeq
In this expression $\Sigma(f)$ is the {\it complexity} of the TAP states with free 
energy density between $f$ and $f+df$.

The existence of two apparently different methods for the study of the static
properties of mean-field spin-glasses clearly poses a problem of
consistency. 
Both these methods are in principle correct, and the equivalence 
between the two approaches should hold in general. A direct argument in its favour is
the cavity approach \cite{MPV,cavity}: on the one hand the cavity approach is
known to be mathematically equivalent to the replica approach; on the other
hand the first step of the cavity approach consists in clustering the
configurations into states which are nothing but the TAP states. From the
cavity analysis, one can expect a priori the existence of a formal
relationship between the replica results computed at a given order $k$ of
replica symmetry breaking, and the results of TAP solutions, correctly
weighted and with an average computed at the order $k-1$. Unfortunately, when
considering the detailed results published so far, there seemed to exist a
discrepancy between the two approaches. For instance the number of TAP
solutions computed directly at the order $k=0$ (annealed average) in
\cite{bm}, disagrees with the one obtained indirectly through the replica
method with one step of replica symmetry breaking (i.e. a $k=1$ computation)
\cite{cavity}. This apparent discrepancy has led us to give a careful look at
this consistency problem. We show here that the TAP approach must be
reconsidered: using a supersymmetry, one finds that the two approaches are
in fact fully consistent.

From a general point of view, the problem of consistency between the TAP approach and
standard thermodynamics  can be posed  in a twofold way.
First, the two methods must agree in the calculation of the
thermodynamic quantities that can be computed in both frameworks. This
requirement is a strong form of consistency. At a weaker level, regarding
objects which are inherently defined only within the TAP method, we may still
expect to find some consistency, and even some formal connections, with the
standard thermodynamic approach.

The strong consistency basically asks that the equilibrium free energy computed 
within the two approaches must be the same. Given that $F_{TAP}$ is a function of the local
magnetization, it is not obvious how to compute the partition function in this framework. More
specifically, due to the large degeneracy of TAP minima, there is the problem of how to weight
correctly these states. This problem was solved by De Dominicis and 
Young \cite{ddy} (DDY), who proposed to weight each TAP state $\alpha$ with,
\beq
w_\alpha = \exp[-\beta F_{TAP}(m^\alpha)] \, / \, Z_{TAP} \ .
\label{pesuzzo}
\eeq
The strong equivalence of the TAP approach to the standard thermodynamic approach
is therefore encoded in the equality (which should hold in the thermodynamic limit)
$\frac{1}{N}\log Z=\frac{1}{N}\log Z_{TAP}$, that is,
\beq
\frac{1}{N}\log \left(\sum_\sigma \exp[-\beta H(\sigma)]\right) =
\frac{1}{N}\log \left(\sum_{\alpha=1}^{\cal N} \exp[-\beta F_{TAP}(m^\alpha)]\right) \ .
\label{strong}
\eeq
Proving relation (\ref{strong}) in general is highly nontrivial, especially because in 
spin-glasses we can only compare averages over the disorder, and thus, from a practical
point of view, we need to compare the average of the replicated partition functions, which
will typically require some tricky integration over complicated order parameters. This program
has been carried out by DDY for the Sherrington-Kirkpatrick (SK) model \cite{sk}. Their conclusion is
that relation (\ref{strong}) is indeed verified, provided that the complexity of the {\it dominant} 
TAP states at any temperature is zero. This restrictive hypothesis is satisfied in the SK model,
but not in other models where the TAP approach is nevertheless used. Thus, it may seem that
either condition (\ref{strong}) is not broadly valid, or that a general proof of it must
give up the hypothesis used by DDY.

In the present paper we show that the proof of (\ref{strong}) given by DDY for the SK model
is valid with no extra hypothesis. The key point is that the TAP partition function can be written 
in an integral form, with an action which is invariant under the Becchi-Rouet-Stora-Tyutin (BRST) 
supersymmetry \cite{brs,tito}.
This symmetry provides the mathematical relations needed to prove
equation (\ref{strong}), without the need to invoke any further hypothesis on the complexity
of the states.

Concerning the weak consistency of the two methods, the key object is the complexity $\Sigma(f)$.
Even though there is no obvious way to compute $\Sigma$ in the standard thermodynamic framework,
we may still impose an important consistency condition: at low temperatures (below the static
transition) we expect the equilibrium thermodynamic states to have the same free
energy density as the lowest TAP states. In other words, we require that at low enough temperatures 
there must be no static contribution of the metastable states. This condition is encoded in the 
equation,
\beq
f_{eq} = f_0 \ ,
\label{weak}
\eeq
where $f_{eq}$ is the equilibrium free energy density computed in the standard thermodynamic framework,
and $f_0$ is defined by the TAP relation $\Sigma(f_0)=0$. 
Apart from this minimal requirement, one may ask whether 
in the TAP calculation of $\Sigma$ emerge some deeper formal connections with the
standard calculation of the free energy. This fact, of course, seems more than likely 
if the strong consistency condition (\ref{strong}) holds, even though it is in 
general not obvious. In the $p$-spin spherical model \cite{crisa1},  
relation (\ref{weak}) is satisfied \cite{kpz,crisatap,monasson,franzparisi}.
Moreover, for this same model it has been shown in \cite{juanpe} that
by means of the BRST supersymmetry the saddle point equations involved in the TAP 
calculation of $\Sigma$ at $f=f_0$ become identical to the static equations in the replica approach. 
In the SK model, however, relation (\ref{weak}) is not easy to check 
\cite{bm,bm1,bm2,dd1,innocent2,potters}, 
since it requires
a full replica symmetry breaking (RSB) calculation of the complexity. Moreover, up to now
there was no formal connection between the TAP complexity and equilibrium free energy 
in the SK model. 

The fact that the BRST supersymmetry plays a crucial role in the $p$-spin,
suggests that even in the SK model this symmetry may help to prove (\ref{weak}),
and discover other formal connections between the two methods. 
This is what we show in Section IV, where we revisit the annealed calculation of the 
SK complexity of Bray and Moore \cite{bm} (BM), and  thanks to the BRST 
supersymmetry show that at $f=f_0$ this calculation becomes equal to the standard 
static calculation of the free energy, at one step level of replica symmetry breaking.
This result comes hardly as a surprise, once the strong consistency (\ref{strong}) of
the two methods is proved. However, there is a subtle point: it is not clear {\it a priori} 
whether a formal connection between static and TAP approaches is valid only for
the correct full RSB solution of the SK model, or if such a connection is preserved at 
each finite (though approximated) level of replica symmetry breaking. Our result
shows that this second scenario is the correct one.

A further, highly nontrivial consistency check of the static and TAP approaches, comes
from a comparison of the TAP complexity with the complexity computed by means of
constrained thermodynamics, which results in the Legendre transform construction of
\cite{monasson}. Within this method the complexity is given by the Legendre 
transform of the thermodynamic free energy of the system, thus establishing a 
connection between TAP and static methods at a {\it generic} value of $f$.
We will prove that, as in the $p$-spin spherical model, in the SK model the TAP
complexity, once the BRST relations are considered, coincides with the complexity 
of \cite{monasson}.

In Section II we introduce the BRST supersymmetry in the context of the TAP approach, and
derive the BRST relations that will be used in the rest of the paper. In Section II we show 
how the BRST relations enforce the DDY proof of (\ref{strong}), without the need of any extra
hypothesis. In Section IV we present the BM calculation of the SK complexity, exploiting the
BRST supersymmetry, and in Section V we show the connection between our results and the Legendre
transform method for the computation of the complexity. Finally, we draw our conclusion in Section VI.

\section{The BRST supersymmetry in the TAP context}

In this Section we show why the BRST supersymmetry is helpful in the context of the 
TAP approach. This is just a specific example of the application of the supersymmetric 
formalism to statistical mechanics, and in particular to the field of disordered systems 
\cite{ps1,ps2,zinnjustin,jorge1,jorge2}. 
As we have seen, the TAP free energy $F_{TAP}$ is a function of the local magnetizations ${m_i}$,
and the minima of this function are identified with the metastable states of the system (TAP states).
In the rhs of equation (\ref{strong}) we have a typical example of a sum over different
TAP states, as often needed in this approach. In addition to the partition function $Z_{TAP}$, we 
may also compute the density of TAP states (\ref{density}),
\beq
\rho(f)=\sum_{\alpha=1}^{\cal N} \delta[F_{TAP}(m^\alpha)-Nf] \ ,
\label{rho}
\eeq
which is the quantity one needs for the computation of the complexity $\Sigma(f)$. 
In (\ref{rho}), as in (\ref{strong}), TAP states are labeled by $\alpha$, and $m^\alpha$
indicates the corresponding set of local magnetizations.
More generally, in the TAP context we always have to deal with expressions of the form,
\beq 
R=\sum_{\alpha=1}^{\cal N} r[F_{TAP}(m^\alpha)] \ ,
\label{suma}
\eeq
$r$ being a generic function of $F_{TAP}$. The quantity $R$ can be written as,
\beq
R=\sum_{\alpha=1}^{\cal N} \int \prod_i dm_i\ \delta(m_i-m_i^\alpha) \ r[F_{TAP}(m)] =
\int \prod_i dm_i\ \delta(\partial_i F_{TAP}(m)) \ |\det (\partial_i \partial_j F_{TAP}(m))| \ r[F_{TAP}(m)]
\ ,
\label{ghirri}
\eeq
where the TAP states have been identified with solutions of the TAP equations $\partial_i F_{TAP}(m)=0$.
In (\ref{ghirri}) the modulus of the determinant is quite hard to handle, and due to this it is 
disregarded in most supersymmetric calculations \cite{discuss}. 
This approximation is {\it a priori} unjustified, since without the modulus 
we are weighting each TAP solution with the sign of the Hessian determinant, with
the risk of uncontrolled cancellations. 
The situation may slightly improve if the function $r(F_{TAP})$ is peaked 
only on low values of the free energy $F_{TAP}$, as it  happens in the calculation of $Z_{TAP}$ at low
temperatures, and of $\rho(f)$ at low free energies. In these cases, we may  hope that at low 
temperatures (free energies) only minima of $F_{TAP}$ will dominate the integral in (\ref{ghirri}). 
Minima have positive-defined Hessian, and thus the modulus becomes redundant. Moreover, the very
identification of TAP states with TAP solution is only sensible if these solutions are minima,
rather than generic saddles.
Yet, the validity of this scenario should be checked carefully, as it can be done for the $p$-spin 
spherical model, where the effect of disregarding the modulus is completely under control, due to the
fact that the free energy distribution of minima and saddles is completely known \cite{mod}.
In the SK model we do not have such a precise information about the TAP free energy saddle points, 
and therefore dropping the modulus is quite risky. However, our aim here is to show that all the
classic TAP calculations (where the modulus was disregarded) are formally consistent with the standard 
static approach, thus justifying {\it a posteriori} this approximation.

We can use an exponential representation for the delta function and the determinant,
\beqa
 \prod_i \delta(\partial_i F_{TAP}) &=& \int_{-i\infty}^{+i\infty} \prod_i \frac{dx_i}{2\pi i}
\ \exp\left(\sum_i x_i \partial_i F_{TAP}(m)\right) \\
\det (\partial_i \partial_j F_{TAP}) &=& \int_{-\infty}^{+\infty} \prod_i d\bar\psi_i\, d\psi_i 
\ \exp\left[\sum_{ij} \bar\psi_i  \psi_j  \partial_i \partial_j F_{TAP}(m) \right] \ ,
\eeqa
where $\{\bar\psi, \psi\}$ are anti-commuting Grassmann variables. In this way we can write \cite{zinnjustin},
\beq
R = \int {\cal D}m\, {\cal D}x \,{\cal D}\bar\psi\, {\cal D}\psi\ \
e^{S(m,x,\bar\psi,\psi)} \ ,
\label{int2}
\eeq
where the action $S$ is given by,
\beq
S(m,x,\bar\psi,\psi)= \sum_i x_i \partial_i F_{TAP}(m) + \sum_{ij}\bar\psi_i  \psi_j  
\partial_i \partial_j F_{TAP}(m) + w[F_{TAP}(m)] \ .
\label{action}
\eeq
In the case $r$ is an ordinary function we have $w=\log r$, while if $r(F_{TAP})=\delta(F_{TAP})$
then $w= u F_{TAP}$, where $u$ is an imaginary integration variable to implement the $\delta$-function (see next 
Section). The measures in (\ref{int2}) include the sum over the indices and the constant prefactors. 
A key property of action (\ref{action}) is its invariance under a generalization of the 
Becchi-Rouet-Stora-Tyutin (BRST) supersymmetry \cite{brs,tito} (see also \cite{zinnjustin}):
if $\epsilon$ is an infinitesimal Grassmann parameter, it is straightforward to verify that 
(\ref{action}) is invariant under the following transformation,
\beq
\delta m_i = \epsilon\, \psi_i \quad\quad
\delta x_i = -\epsilon \, w^\prime \, \psi_i \quad\quad
\delta \bar\psi_i = -\epsilon\, x_i \quad\quad
\delta \psi_i = 0 \quad\quad \quad \Rightarrow \quad\quad \delta S=0
\ .
\label{brst}
\eeq
This  generalization of the standard BRST supersymmetry to the case where $w\neq 0$ has been first 
introduced in \cite{juanpe}, in the context of a TAP calculation for the $p$-spin spherical model.
The BRST invariance does not depend on the explicit form of the function $F_{TAP}(m)$, but simply on the
formal structure of action (\ref{action}). In particular, it is essential that $w$ is a function of the
magnetizations $m_i$ only through $F_{TAP}(m)$. The fact that $\delta S=0$ under the BRST supersymmetry
implies that the average of any observable of the same variables performed with this action must be 
invariant too. For an observable ${\cal O}$,
\beq
\langle {\cal O}(\Lambda) \rangle =\int \ {\cal D} \Lambda \ {\cal O}(\Lambda) \ e^{S(\Lambda)} \ ,
\label{gaga}
\eeq
with $\Lambda=\{m,x,\bar\psi,\psi\}$, we have
\beq
\langle{\cal O}(\Lambda) \rangle -\langle {\cal O}(\Lambda-\delta\Lambda) \rangle=
\langle \delta {\cal O}(\Lambda) \rangle=0 \ .
\label{gugu}
\eeq 
This property can be used to generate some useful Ward identities. 
A fruitful choice is ${\cal O}= m_i^k \bar\psi_i$ and ${\cal O}= x_i^k\bar\psi_i$, whose variation gives,
\beqa
k\,\langle m_i^{k-1}\bar\psi_i \psi_i \rangle &=& - \langle m_i^{k} x_i \rangle 
\label{brsa}\\
k\,\langle  x_i^{k-1} w^\prime \bar\psi_i \psi_i \rangle &=&\phantom{-} \langle x_i^k x_i \rangle \ .
\label{brsb}
\eeqa
In particular, the case $k=1$ gives the BRST equations already used in \cite{juanpe},
\beqa
\langle \bar\psi_i \psi_i \rangle &=& - \langle m_i x_i \rangle  \quad\quad {\rm [BRST 1]}
\label{brst1}
\\
\langle w^\prime \bar\psi_i \psi_i \rangle &=&\phantom{-} \langle x_i x_i \rangle \quad\quad \ {\rm [BRST 2]} \ .
\label{brst2}
\eeqa
The BRST relations are crucial. Thanks to them we will be able to prove in general that any TAP average 
must give the same result as the equivalent thermodynamic average, and more specifically to reduce the 
TAP calculation of the complexity to the standard replica calculation of the free energy.

\section{The De Dominicis-Young calculation revisited}

In 1983, De Dominicis and Young \cite{ddy} (DDY) introduced the statistical weight (\ref{pesuzzo}) 
for the TAP states and explicitly proved that {\it under a key hypothesis} the average of
the replicated TAP partition function is equal to the average of the replicated standard 
partition function, i.e.
\beq
\int {\cal D}J\, P(J) Z^n(J)=\int {\cal D}J\, P(J) Z_{TAP}^n(J) \ ,
\eeq
which is the practical way to write equation (\ref{strong}). 
This result is very important, because it proves
the identity of the TAP and static approaches {\it before} any Ansatz for the overlap matrix
is done. The hypothesis that DDY invoked to justify their computation is that 
the complexity of the dominant  TAP states at any temperature is zero. 
More precisely, we can write,
\beq
Z_{TAP}=\int df\ \rho(f)\ \exp[{-\beta N f}]=
\int df\  \exp[-\beta N( f-T\Sigma(f))]=  \exp[-\beta N( f^\star-T\Sigma(f^\star))] \ ,
\label{monaco}
\eeq
where the dominant free energy $f^\star$ is solution of the equation,
\beq
\beta=\frac{\partial\Sigma(f^\star)}{\partial f} \ .
\eeq
DDY assumed as a necessary condition that
\beq
\Sigma(f^\star)=0 \ .
\label{no}
\eeq
In the present Section we argue that, although relation (\ref{no}) is in fact
verified in the SK model, it is {\it not} a necessary condition for the self-consistency of 
the TAP approach and for its equivalence with the statics. 

\subsection{The problem and its solution in the DDY formulation}

Let us briefly summarize the arguments of DDY leading to condition (\ref{no}) (in order to 
keep the notation as clear as possible we treat $m$ as a one-dimensional variable in 
this Section; things do not change with the full $N$-dimensional representation). 
First, we introduce an auxiliary magnetic field $h$ both in $F_{TAP}(m)$
through the term $-hm$, and in the Hamiltonian $H(\sigma)$ with $-h\sigma$. 
From the very definition of equilibrium magnetization and energy we trivially have,
\beqa
\frac{1}{\beta} \, \frac{\partial Z}{\partial h} &=& \sum_\sigma \sigma \,
\exp[-\beta H(\sigma)]=
\langle \sigma\rangle \, Z\ ,
\label{campo} \\
- \frac{\partial Z}{\partial \beta} &=&  \sum_\sigma H(\sigma)\,  
\exp[-\beta H(\sigma)]=
\langle H \rangle \, Z\ .
\label{energia}
\eeqa
On the other hand,  from (\ref{strong}) we get,
\beqa
\frac{1}{\beta} \, \frac{\partial Z_{TAP}}{\partial h} &=& X -\sum_\alpha^{{\cal N}(h,\beta)} 
\frac{\partial F_{TAP}(m_\alpha)}{\partial h}\ \exp[-\beta F_{TAP}(m_\alpha)]  =
X + \langle\sigma\rangle\, Z 
\label{xx}
\\
-\frac{\partial Z_{TAP}}{\partial \beta} &=& Y+ \sum_\alpha^{{\cal N}(h,\beta)} 
\frac{\partial F_{TAP}(m_\alpha)}{\partial \beta}\ \exp[-\beta F_{TAP}(m_\alpha)]  =
Y + \langle H \rangle\, Z  \ .
\label{xy}
\eeqa
To obtain this result we used the following relation,
\beq 
-\frac{\partial F_{TAP}(m_\alpha)}{\partial h} = 
\sum_{\sigma\in\alpha} \sigma \, \exp[-\beta H(\sigma)] \ ,
\eeq
where the sum over $\sigma$ is restricted to those configurations belonging
to state $\alpha$ (an analogous relation is valid for $\partial F_{TAP}/\partial\beta$). 
On the other hand, in equations (\ref{xx})-(\ref{xy}) $X$ and $Y$ are the 
contributions coming, respectively, from the dependence on $h$ and $\beta$ 
of the total number of TAP states, ${\cal N}(h,\beta)$. 
DDY correctly noted that $X=0$ and $Y=0$ are necessary conditions for
the consistency of the TAP approach.
Moreover, by using (\ref{monaco}), they noted that
if $\Sigma(f^\star)=0$, these extra contributions are indeed vanishing in the thermodynamic 
limit, and thus consistency is recovered. Thus, DDY assumed  $\Sigma(f^\star)=0$ and 
accordingly imposed $X=Y=0$ in their computation. More precisely, after 
averaging over the disorder, the relations $X=Y=0$ become two equations for 
the auxiliary fields (eqs.(32) and (42) of \cite{ddy}), which DDY used to prove (\ref{strong}).
Our point, however, is that the equations $X=Y=0$ are always satisfied, 
irrespective of the value of $\Sigma(f^\star)$. 

\subsection{Role of the supersymmetry}

We now show that the mathematical conditions 
on the auxiliary variables that DDY impose are nothing else than the BRST relations, which are 
valid in general, and contain no information on $\Sigma(f^\star)$.
Let us define,
\beq
G(m,h)=\frac{\partial (\beta F_{TAP})}{\partial m} \quad\quad , \quad\quad r(m,h)=e^{-\beta F_{TAP}(m,h)} \ .
\eeq
We consider relation (\ref{campo}) (a similar reasoning can be easily done also for 
relation (\ref{energia})) and use the integral representation of Section II for $Z_{TAP}$, 
to obtain,
\beq
\frac{\partial Z_{TAP}}{\partial h}=
-\int dm\ dx \ x\, \,e^{x G(m,h)} \, \frac{\partial G}{\partial m}\ r(m,h) +
\int dm\ dx \ e^{x G(m,h)} \, \frac{\partial G}{\partial m}\ \frac{\partial r}{\partial h} \ .
\label{ruminante}
\eeq
In this expression the first term is $X$, whereas the second term is just $\langle \sigma\rangle \, Z$.
Integrating by parts we have,
\beq
X=  \int dm\ \delta(G(m,h)) \ \frac{\partial r}{\partial m} \ .
\eeq
At this point note that for a generic function $r(m,h)$ there is no need for $X$ to be zero, in much
the same way as action (\ref{action}) is not BRST invariant if $r$ is not a function of $F_{TAP}$.
If, however, $r(h,m)=r(F_{TAP}(m,h))$, as in the case under examination, we have,
\beq
X= \int dm\ \delta(G(m,h)) \ G(m,h)\ \frac{\partial r}{\partial F_{TAP}} \ =0 \ .
\label{mispiego}
\eeq
Thus, the extra term coming from the differentiation of ${\cal N}(h,\beta)$ with respect to $h$
vanishes if the TAP states are weighted with a function of the TAP free energy. A similar 
conclusion can be easily drawn for differentiation with respect to $\beta$, which leads to
$Y=0$. In particular, these relations hold in the case $r(F_{TAP})=\exp(-\beta F_{TAP})$, which
is the one analyzed by DDY.

The important fact is that from a mathematical point of view, $X=0 \ ,Y=0$ are 
consequences of the particular form of action  (\ref{action}), and of its symmetry 
properties. Indeed, if we express (\ref{campo}) and (\ref{energia}) in the supersymmetric 
representation (\ref{gaga}), we get,
\beqa
X &=&  \langle x \rangle \ \non \\
Y &=&  \langle \left[  x + \bar\psi \psi \frac{\partial}{\partial m}\right]
\frac{\partial G}{\partial\beta} \rangle = \sum_k c_k \langle 
 m^{k} x\rangle  + k \ \langle m^{k-1} \bar\psi \psi \rangle \ ,
\label{dudu}
\eeqa
with,
\beq 
c_k= \frac{1}{k!}\ \frac{ \partial^k}{\partial m^k} \ \frac{\partial G}{\partial \beta} \ .  
\eeq
A comparison between (\ref{dudu}) and (\ref{brsa}) clearly shows that the relations $X=0$ and $Y=0$ are 
a direct consequence of the BRST symmetry.  Thus, the relations  imposed by DDY in 
the calculation of $Z_{TAP}^n$, and in their demonstration of the equivalence between 
TAP and static averages, are a consequence of the BRST supersymmetric form of the TAP 
action (\ref{action}). 
These relations are therefore not the expression for the absence of an extensive complexity
of the equilibrium states, i.e. they do not imply $\Sigma(f^\star)=0$. For this reason,
the result of DDY is more general than what originally thought. 
This is consistent with the fact that a similar formal connection between TAP and static 
approaches is valid in the $p$-spin spherical
model, where, in a certain range of temperature, $\Sigma(f^\star) > 0$.

\subsection{Disregarding the modulus}

Before closing this Section, a word of caution on the modulus of the determinant
\cite{ps2,jorge1,mod}. As already noted
above, disregarding the modulus can be very risky, especially if the function $r(F_{TAP})$ weighting
the TAP states in (\ref{suma}) is not peaked on low free energies, and thus minima. 
As an extreme illustration of this risk we consider the function $r(F_{TAP})=1$. 
Clearly, from (\ref{suma}) we have,
\beq
R(h)={\cal N}(h) \quad \quad \Rightarrow \quad\quad \frac{d R}{d h}=\frac{d{\cal N}}{d h} \neq 0 \ .
\label{billo}
\eeq
However, if we use for $R$ the integral representation (\ref{ghirri}) {\it and} we disregard the 
modulus, we obtain,
\beq
\frac{d R}{d h}=0 \ ,
\label{morse}
\eeq
since the BRST supersymmetry is trivially satisfied with $r(F_{TAP})=1$. The problem here
is that by disregarding the modulus, and with a flat weight $r$, we are summing over {\it all} stationary 
points of $F_{TAP}$, each multiplied by the sign of the determinant. The Morse theorem states that this quantity
must be a topological constant, only dependent on the manifold over which $F_{TAP}$ is define and on the
boundary conditions on $F_{TAP}$. Result 
(\ref{morse}) is therefore correct, but the quantity in this equation is not the same $R(h)$ as in (\ref{billo}). 
This trivial example shows how important is to weight TAP solutions with a function peaked as much as possible on
low free energies when the modulus is disregarded.

\section{The Bray-Moore calculation revisited}

In the present Section we will compute the annealed complexity of the TAP states for the
SK model, following closely the classic calculation of Bray and Moore (BM) \cite{bm}.
Our new contribution will be to exploit the BRST relations (\ref{brst1}) and (\ref{brst2}) 
in order to simplify the resulting saddle point equations. In this way we will prove that 
the complexity is intimately connected to the 1RSB static free energy.

\subsection{General definitions}

The TAP free energy for the SK model is given by \cite{tap},
\beq
F_{TAP}(m)=-\frac{1}{2} \sum_{ij} J_{ij} m_i m_j + \frac{1}{\beta} \sum_i \phi_0(q,m_i) \ ,
\label{ftap}
\eeq
with,
\beqa
\phi_0(q,m)&=&
\frac{1}{2}(1+m)\, \log\left[\frac{1}{2}(1+m)\right] +
\frac{1}{2}(1-m)\, \log\left[\frac{1}{2}(1-m)\right] -\frac{\beta^2}{4}(1-q)^2
\non
\\
&=& \frac{1}{2}\log(1-m^2) + m\,\tanh^{-1}(m) -\log 2 -\frac{\beta^2}{4}(1-q)^2 \ .
\label{phi0}
\eeqa
The variables $m_i$ are the local magnetizations, and $q$ is  the self-overlap of the TAP states,
\beq
q=\frac{1}{N}\sum_i m_i^2 \ ,
\eeq
while the quenched couplings $J$ are random variables with Gaussian distribution,
\[
P(J_{ij})=\sqrt{N/2\pi}\ \exp(-NJ_{ij}^2/2) \ .
\non
\]
The TAP equations and the Hessian of the free energy are respectively,
\beqa
\beta \, \partial_i F_{TAP}(m) &=& -\beta \sum_{j\neq i} J_{ij} m_j + \phi_1(q,m_i) = 0 \ , \non\\
\beta \, \partial_i \partial_j F_{TAP}(m) &=& -\beta J_{ij} + \phi_2(q,m_i) \, \delta_{ij} \non \ ,
\eeqa
with,
\beqa
\phi_1(q,m) &=& \beta^2 (1-q) m + \tanh^{-1}(m) \non \\
\phi_2(q,m) &=& \beta^2 (1-q) + \frac{1}{1-m^2} \ + \,O(1/N) \ .
\label{phi}
\eeqa
The term of order $1/N$ in $\phi_2(q,m)$ will be dropped in what follows. 
Following BM we perform an {\it annealed} calculation of the number of TAP states, 
i.e. we directly average $\rho(f,\beta|J)$ in (\ref{rho})
over the distribution of the quenched couplings $J_{ij}$,
\[
{\cal N}(\beta,f)=\int {\cal D}J \, P(J) \, \rho(f,\beta|J) \ .
\non
\]
We use for $\rho$ the integral representation of eqs. (\ref{int2}) and
(\ref{action}), with 
\beq
\delta(F_{TAP}-N\, f)  = \int_{-i\infty}^{+i\infty} \frac{du}{2\pi i} 
\ \exp[\,u\,(F_{TAP}-Nf)] \ ,
\eeq
and thus $w(F_{TAP})=u\, (F_{TAP}-Nf)$ and $w^\prime=u$. Thus, the average number of TAP states becomes,
\beq
{\cal N}(\beta,f)= \int {\cal D}J \, P(J) \ {\cal D}m\, {\cal D}x \,{\cal D}\bar\psi\, {\cal D}\psi\, du \
e^{\beta S(m,x,\bar\psi,\psi,u)} \ ,
\label{pino}
\eeq
where we have multiplied the action by $\beta$ in order to keep our calculation as close as possible
to BM.
Before proceeding we mention an important point: in the calculation of BM the $J$-dependent 
part of $F_{TAP}(m)$ in $S$  is eliminated by using the equations $\partial_i F_{TAP}(m)=0$, 
which are enforced by the $\delta$-function. More specifically, BM use the equation,
\beq
-\frac{1}{2} \sum_{ij} J_{ij} m_i m_j=-\frac{1}{2\beta } \sum_{i} m_i \phi_1(q,m_i) \ ,
\eeq
which is valid in the TAP states.
This substitution simplifies considerably the calculation, but unfortunately the action obtained in this way
is no longer BRST invariant. Thus, in the present calculation we must use the full form of $F_{TAP}(m)$,
equation (\ref{ftap}), and for this reason some parts of the calculation differ from BM.

\subsection{The calculation}

The $J$-dependent part of the action in (\ref{pino}) is given by,
\beq
\beta S_J= -\beta\sum_{ij} J_{ij} (x_i m_j + \bar\psi_i\psi_j + \frac{1}{2}u \ m_i m_j ) \ ,
\eeq
and after averaging $\cal N$ over the disorder we obtain the new effective action,
\beqa
\beta S &=& \sum_i x_i \phi_1(q,m_i) + \sum_{i}\bar\psi_i\psi_i \phi_2(q,m_i) + 
u \sum_i  \phi_0(q,m_i) - N \beta u f \non \\
&+& \frac{\beta^2 q}{2} \sum_i x_i^2 + \frac{\beta^2}{2N}\sum_i(m_ix_i)^2
-\frac{\beta^2}{2N}\sum_i(\bar\psi_i\psi_i)^2 + N\frac{\beta^2}{4}u^2 q^2 + \beta^2 u q\sum_i m_i x_i \ .
\label{effective}
\eeqa
In order to linearize the quadratic terms, we introduce in (\ref{pino}) the following $\delta$-functions,
\beqa
\delta(qN-\sum_i m_i^2) &=&\int_{-i\infty}^{+i\infty} d\lambda \ e^{-\lambda q N + \lambda\sum_i m_i^2} \\
\delta(RN-\sum_i m_i x_i)&=&\int_{-i\infty}^{+i\infty} dr \ e^{-r R N + r\sum_i m_i x_i} \\
\delta(TN-\sum_i \bar\psi_i\psi_i)&=&\int_{-i\infty}^{+i\infty} dt \ e^{-t T N + t\sum_i \bar\psi_i\psi_i} \ ,
\label{deltas}
\eeqa
and we integrate over $q,R$ and $T$. In this way the integrals in $x_i$ and $(\bar\psi_i,\psi_i)$ become
Gaussian and can be performed explicitly. The effective action becomes,
\beq
\beta S(\Omega,m)= N\Sigma_0(\Omega) +\sum_i {\cal L}(\Omega,m_i) \ ,
\eeq
where $\Omega=\{q,\lambda,r,R,t,T,u\}$, and,
\beqa
\Sigma_0(\Omega)&=&\frac{\beta^2}{2}R^2 - \frac{\beta^2}{2}T^2 - r R - t T -\frac{1}{2}\log(2\pi\beta^2 q)
-u f \beta - \lambda q  + \frac{\beta^2}{4}u^2q^2 + \beta^2 u q R \ , \non \\
{\cal L}(\Omega, m)&=&u\phi_0(q,m)-\frac{1}{2\beta^2 q}[\phi_1(q,m)+rm]^2+\log[\phi_2(q,m)+t]  + \lambda m^2  \ .
\label{pezzi}
\eeqa
We can now write the average number of TAP states as,
\beq
\overline{{\cal N}(\beta,f)}=\int {\cal D}\Omega \ e^{N\Sigma_0(\Omega)}\ \prod_i\int dm_i 
\ e^{{\cal L}(\Omega,m_i)}=
\int {\cal D}\Omega \ e^{N\Sigma_0(\Omega) + N \log\int dm \ e^{{\cal L}(\Omega,m)}} \ .
\eeq
Thanks to the prefactor $N$ in the exponential, the 
integral in ${\cal D}\Omega$ can be performed with the steepest descent method.
In this way we can write the complexity $\Sigma(\beta,f)$ as,
\beq
\Sigma(\beta,f)=\frac{1}{N}\log \overline{{\cal N}(\beta,f)}= \Sigma_0(\hat\Omega) + 
\log\int dm \ e^{{\cal L}(\hat\Omega,m)}
\ ,
\label{complex}
\eeq
where $\hat\Omega$ is solution of the saddle point equations,
\beq
\frac{\partial \Sigma_0(\Omega)}{\partial\Omega} \ + 
\ \langle\langle \frac{\partial {\cal L}(\Omega,m) }{\partial\Omega} \rangle\rangle =0 \ ,
\label{saddle}
\eeq
with,
\beq
\langle\langle {\cal O}(m) \rangle\rangle = \
\frac{1}{\int dm \ e^{{\cal L}(\Omega,m)}}\ 
\int dm \  {\cal O}(m) \ e^{{\cal L}(\Omega,m)}  \ . 
\eeq
We can reduce the number of variables by directly solving the saddle point equations for $R$ and $T$,
\beqa
\frac{\partial\Sigma_0}{\partial R}&=&0 \quad\Rightarrow\quad  R=r/\beta^2-q\, u \non \\
\frac{\partial\Sigma_0}{\partial T}&=&0 \quad\Rightarrow\quad T=-t/\beta^2  \ .
\label{partial}
\eeqa
In order to have expressions as similar as possible to the ones of BM, we define,
\beqa
B&=&\beta^2(1-q)+t \non \\
\Delta&=&-\beta^2(1-q)-s \ .
\label{defa}
\eeqa
Using the explicit forms of $\phi_1, \phi_2$ in (\ref{phi}), and relations (\ref{partial}) and
(\ref{defa}), we can rewrite ({\ref{pezzi}) as,
\beqa
\Sigma_0(\Omega) &=& -\lambda q - \beta u f - (B+\Delta)(1-q) + \frac{(B^2-\Delta^2)}{2\beta^2} \non
\\
&&\phantom{guuuuuuuuuus}
-\frac{1}{2}\log(2\pi\beta^2q) - \frac{\beta^2}{4}u^2q^2 -u q\Delta - \beta^2 u q(1-q) \ ,
\label{sigma} \\
{\cal L}(\Omega, m)&=& \log\left(\frac{1}{1-m^2}+B\right) - 
\frac{[\tanh^{-1}(m)-\Delta m]^2}{2\beta^2q} + \lambda m^2 + u \phi_0(q,m) \ ,
\label{elle}
\eeqa
where $\phi_0$ is given in (\ref{phi0}). The remaining saddle point equations (\ref{saddle}) for
the variable $\{\lambda,u,B,\Delta,q\}$ are,
\beqa
\frac{\partial \Sigma}{\partial \lambda}\,=0 
\quad &\Rightarrow& \quad\quad
q=\langle\langle m^2 \rangle\rangle 
\label{sad1}
\\
\frac{\partial \Sigma}{\partial u} \,=0
\quad &\Rightarrow& \quad\quad
\beta f = \langle\langle \phi_0(q,m) \rangle\rangle -\frac{\beta^2}{2}u q^2 -q\Delta -\beta^2 q(1-q)
\label{sad2}
\\
\frac{\partial \Sigma}{\partial B} \,=0
\quad &\Rightarrow& \quad\quad
B\left[ 1-\beta^2 \langle\langle \frac{(1-m^2)^2}{1+B(1-m^2)} \rangle\rangle \right] =0
\label{sad3}
\\
\frac{\partial \Sigma}{\partial \Delta} \,=0
\quad &\Rightarrow& \quad\quad
\Delta=-\frac{\beta^2}{2}(1-q) + \frac{1}{2q}\langle\langle m\tanh^{-1}\, (m)\rangle\rangle
-\frac{\beta^2}{2}u q
\label{sad4}
\\
\frac{\partial \Sigma}{\partial q} \,=0
\quad &\Rightarrow& \quad\quad
\lambda= B + \Delta - \frac{1}{2q}\left\{ 1 - \frac{1}{\beta^2q}
\langle\langle\left[ \tanh^{-1}\, (m)-\Delta m\right]^2\rangle\rangle \right\} \non
\\
&& \phantom{buuuuuuuuuuuuuuuuuuuuuu}
+ u\langle\langle \frac{\partial \phi_0(q,m)}{\partial q} \rangle\rangle
-\left[ \frac{\beta^2}{2}u^2 q +u\Delta + \beta^2 u (1-2q) \right] \ .
\label{sad5}
\eeqa
These equations, together with relations  (\ref{sigma}) and (\ref{elle}) 
can be compared with equations (15), (16) and (17) of BM \cite{bm}. 
The differences are due to the different representation of $F_{TAP}(m)$ we have 
taken, in order to preserve the BRST supersymmetry. However, the final result, i.e. 
the values of $\Sigma_0$ and $\cal L$ in the solution of the saddle point
equations, is exactly the same (in the last of equations (17) of BM, 
however, there is a term $u\,\partial f/\partial q$ missing). Of course,
if we set $u=0$, i.e. if we do not impose the constraint on the free energy, our
expressions become formally identical to BM.

\subsection{Role of the supersymmetry and connection with the statics}

Solving the saddle point equations above, even numerically, is not a simple task.
However, the problem becomes much easier if we make use of the
two BRST relations (\ref{brst1}) and (\ref{brst2}).
From (\ref{deltas}), (\ref{partial}) and (\ref{defa}), we have,
\beqa
\langle \bar\psi_i\psi_i\rangle &=& T = -t/\beta^2= -B/\beta^2 + (1-q) \non \\
\langle x_i m_i\rangle &=& R = r/\beta^2 - q\, u= -\Delta/\beta^2 - (1-q) - q\,  u \ ,\non 
\eeqa
and thus the first BRST equation becomes,
\beq
\phantom{oooooooooooooooooooooi}
B+\Delta= - \beta^2 q\, u \phantom{oooooooooooooooooooooi}  {\rm [BRST1]} \ .
\label{brst11}
\eeq
In order to write the second BRST equation we need $\langle x_i x_i\rangle$, which can be 
computed from equation (\ref{effective}),
\[
\langle x_i x_i \rangle = 
-\frac{1}{\beta^2q}
\left\{ 1 - \frac{1}{\beta^2 q}\langle\langle [\tanh^{-1}(m)-\Delta m]^2 \rangle\rangle\right\} \ .
\]
In this way the second BRST relation becomes (we recall that $w^\prime=u$),
\beq
\frac{1}{q}\left\{ 1 - \frac{1}{\beta^2 q} \langle\langle [\tanh^{-1}(m)-\Delta m]^2 \rangle\rangle
\right\}= u[B-\beta^2(1-q)] \quad\quad\quad\quad {\rm [BRST2]} \ .
\label{brst22}
\eeq
Equation (\ref{sad3}) admits the solution $B=0$, and following BM this is the solution we adopt.
Setting $B=0$ into the first BRST relation we obtain,
\beq
\Delta=-\beta^2q u \label{brst111} \ , 
\eeq
while substituting equation (\ref{phi0}) and the second BRST relation into (\ref{sad5}), we find,
\beq
\lambda=\frac{1}{2} \beta^2 u^2 q \label{brst222} \ .
\eeq
We now use these two results, eqs. (\ref{brst111}) and (\ref{brst222}), 
to rewrite the complexity $\Sigma(\beta,f)$ in (\ref{complex}). 
By making the change of variable $m\to h=\tanh^{-1}(m)$
in the integral in (\ref{complex}), and using (\ref{sigma}) and (\ref{elle}), we obtain,
\beqa
\Sigma(\beta,f) = u\left\{ -\log 2 -\frac{1}{2u}\log(2\pi\beta^2q) + \frac{\beta^2}{4}
[(-u-1)q^2+2q-1] + \frac{1}{u}\log\int dh \ e^{{\cal F}(h;q,u)} \right\}
-\beta u f \ ,
\label{apple}
\eeqa
with,
\beq
{\cal F}(h;q,u)=-\frac{h^2}{2\beta^2q}-u\log\cosh h \ .
\label{effe}
\eeq
This form of the complexity can be fully appreciated by recalling the expression for the free 
energy of the SK model at the 1-step level of replica symmetry breaking (1RSB). The 1RSB free energy 
density is given by \cite{1rsb},
\beq
\beta F_{1RSB}(\beta;q_1,x)= -\log 2 + \frac{1}{2x}\log(2\pi\beta^2q_1) + \frac{\beta^2}{4}
[(x-1)q_1^2+2q_1-1] - \frac{1}{x}\log\int dh \ e^{{\cal F}(h;q_1,-x)} \ ,
\eeq
where the self-overlap $q_1$ and the breaking point $x$ satisfy the saddle point equations 
$\partial F_{1RSB}/ \partial q_1=0$,  $\partial F_{1RSB}/ \partial x=0$,  and where we have set 
to zero the mutual overlap $q_0$. 
Clearly, we see that there is a striking formal correspondence between
$\Sigma$ and $F_{1RSB}$. Indeed we have,
\beq
\Sigma(\beta,f;q,u)= \beta u\,[F_{1RSB}(\beta;q,-u)-f] \ ,
\label{end}
\eeq 
and,
\beqa 
0= \frac{\partial\Sigma}{\partial q}&=&\beta u \frac{\partial F_{1RSB}(\beta;q,-u)}{\partial q} 
\label{parz1}
\\
0= \frac{\partial\Sigma}{\partial u}&=&\beta u \frac{\partial F_{1RSB}(\beta;q,-u)}{\partial u} +
\beta [F_{1RSB}(\beta;q,-u)-f] \ .
\label{parz2}
\eeqa
Therefore the saddle point equations for $\Sigma$ and $F_{1RSB}$ 
coincide, provided that $f=F_{1RSB}(\beta;q_1,x)$. 
Moreover, for $f=F_{1RSB}$ we trivially  have
\beq
\Sigma[\beta,F_{1RSB}(\beta;q_1,x)]=0 \ ,
\eeq
and therefore $f_0=F_{1RSB}$: as expected, and anticipated in the 
introduction, the lowest TAP states have the static free energy density.

The fact that $q_1=q$ and $x=-u$, 
at $f_0$ can also be obtained by directly setting $f=f_0=F_{1RSB}$
in the saddle point equations (\ref{sad1}-\ref{sad5}).
From (\ref{sad2}) and (\ref{sad4}), after some algebra, we obtain,
\beq
-\frac{1}{4}\beta^2q^2u+\frac{1}{2u}\log(2\pi\beta^2q) -\frac{1}{u}\log\int dh \ e^{{\cal F}(h;q,u)}
-\langle\langle \log\cosh h\rangle\rangle=0 \ , \label{supersad1}
\eeq
while equation (\ref{sad1}) becomes,
\beq
1+\beta^2\,q\,u\,[(u+1)q-1] -\frac{1}{\beta^2 q}\langle\langle h^2 \rangle\rangle = 0 \ ,
\label{supersad2}
\eeq
where the averages $\langle\langle\cdot\rangle\rangle$ are now performed with the
distribution $\exp[{\cal F}(h;q,u)]$ of (\ref{effe}). Once we set $u=-x$ and $q=q_1$ equations 
(\ref{supersad1}) and (\ref{supersad2}) coincide with the 1RSB saddle point equations, i.e.
$\partial F_{1RSB}/\partial x=0$ and $\partial F_{1RSB}/\partial q_1=0$ \cite{1rsb}. 
Summarizing, the complexity $\Sigma$ vanishes at the 1RSB free energy
density, and at this point the two calculations are formally equivalent, since the 
saddle point parameters coincide, and $\Sigma$ and $F_{1RSB}$ are  related by 
equation (\ref{end}). 

Some considerations are in order. First, this identification of the saddle point 
parameters in the two calculations has a clear physical meaning, already discussed 
in \cite{juanpe}. The self-overlap
$q_1$ of the 1RSB states is just the same as the self-overlap of the TAP states
at $f=F_{1RSB}$. Less obvious is the relation $u=-x$, since the 1RSB breaking point 
$x$ does not seem trivially related to the parameter $u$. However, we must note that,
\beq
\frac{d\Sigma(f)}{d f}=-\beta u(f) \ ,
\eeq
and thus $-\beta u(F_{1RSB})$ is the slope of the complexity at the lowest free energy, which
is indeed equal to the static breaking point $\beta x$.

Secondly, we recall that we have set $q_0=0$ in the 1RSB calculation. This is due to the 
fact that we performed an annealed computation of $\Sigma$, and thus we only had one value
of the overlap. We believe that a quenched, but replica symmetric, calculation of
$\Sigma$ will be equivalent to the 1RSB static calculation with $q_0\neq 0 $. This brings us to 
the final point: the TAP method has
one less step of replica symmetry breaking as compared to the standard static one. This
is simply due to the fact that the elementary objects in the TAP approach are states, 
while in the static one are configurations. For this reason, a model as the $p$-spin spherical
spin glass, which is exactly solved by a 1RSB Ansatz, has a complexity which is exactly
replica symmetric. In the SK model, however, which needs a full RSB static solution, also the 
correct complexity will need to be computed at a full RSB level \cite{juanpe}.

\subsection{Numerical solution of the saddle point equations}

We now consider the saddle point equations for $\Sigma$ at a generic value of $f\ge F_{1RSB}$. We are left
with two unknown variable, $q$ and $u$, and with three unused equations, that is (\ref{sad1}), 
(\ref{sad2}) and (\ref{sad4}). However, by using the BRST relations it is possible to show that 
actually equations (\ref{sad1}) and (\ref{sad4}) coincide. To prove this fact one must use the formula,
\beq
\langle\langle h^2 \rangle\rangle = \beta^2 q -\beta^4q^2u + \beta^4q^2 u(u+1) 
\ \langle\langle \tanh^2 h \rangle\rangle \ ,
\eeq
in (\ref{brst22}), and then substitute this second BRST relation into either (\ref{sad1}), 
or (\ref{sad4}). Thus, the two remaining saddle point equations for $q$ and $u$, (\ref{sad1}) and 
(\ref{sad2}), become,
\beqa
q &=&\langle\langle \tanh^2 h \rangle\rangle \non \\
\beta f &+& \langle\langle \log\cosh h \rangle\rangle + 
\frac{\beta^2}{4}\, [q^2(1+2u) -2q+1] + \log 2 =0  \ .
\label{menose}
\eeqa
where the formula,
\[
\frac{1}{2}\log(1-m^2) = -\log\cosh h \ , 
\]
was used.
\begin{figure}
\includegraphics[clip,width=4 in]{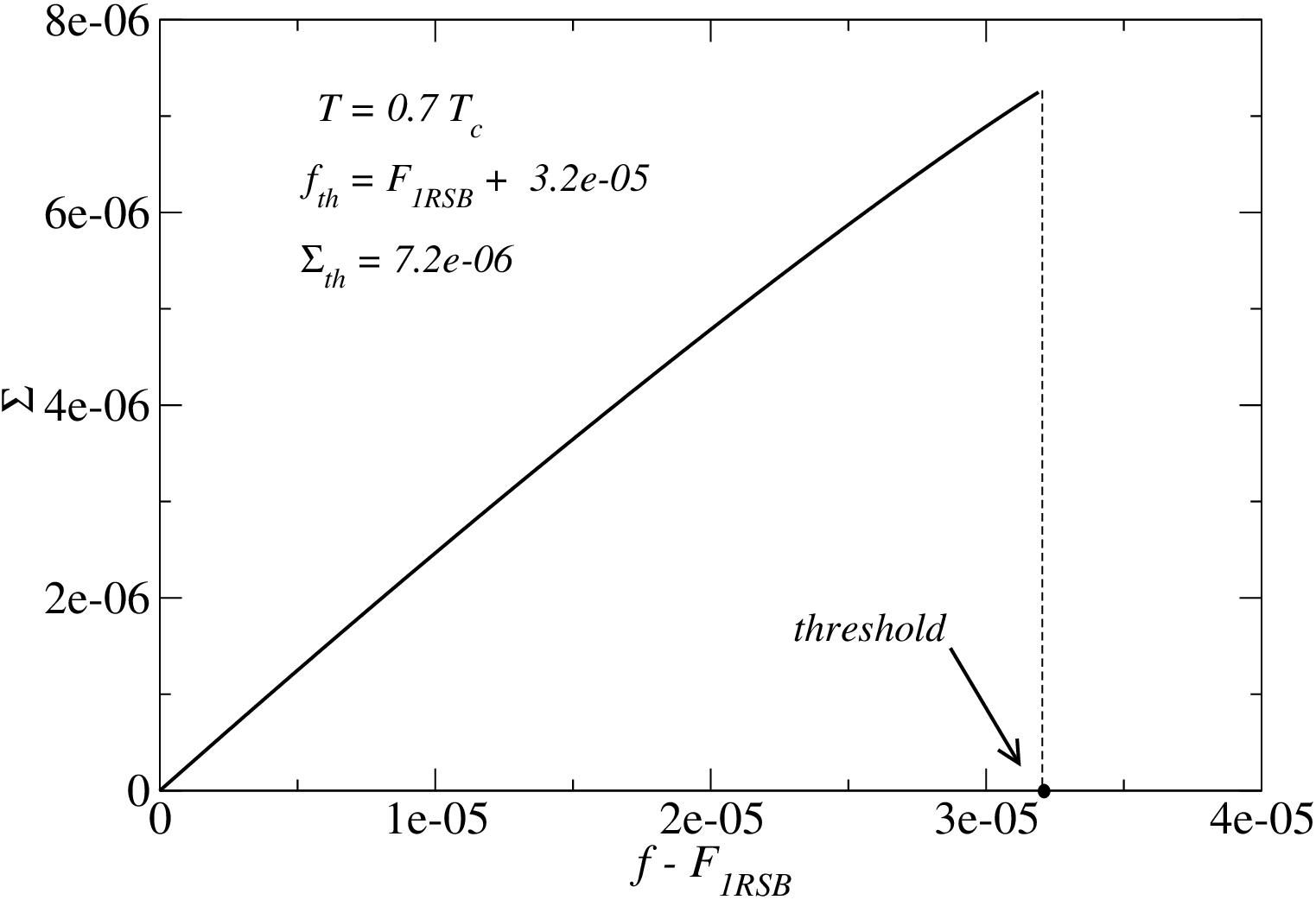}
\caption{Complexity $\Sigma$ as a function of the free energy density difference $f-F_{1RSB}(\beta)$, for
$T =0.7 \, T_c$.}
\label{sigmafig}
\end{figure}
It is not difficult to solve numerically the two equations above, and this shows how
drastic is the simplification of the calculation due to the use of the BRST 
supersymmetry. In Fig.1 we plot $\Sigma$ as a function of $f-F_{1RSB}(\beta)$, 
at $T=0.7\ T_c$ (we recall that $T_c=1$). The system of equations (\ref{menose}) 
have solution only up to 
a threshold value $f=f_{th}(\beta)$, where the complexity takes its maximum value 
$\Sigma_{th}$. The threshold free energy turns out to be quite close to 
$F_{1RSB}(\beta)$. An expansion of the equations close to $T_c$ shows that 
$f_{th}=F_{1RSB}+O(\epsilon^5)$, where $\epsilon=T_c-T$, whereas $q=O(\epsilon)$
and $u=O(\epsilon)$.

An interesting comparison with the result of BM can be done by computing
the {\it total} complexity of TAP solutions, $\Sigma_{tot}$. BM obtain this
quantity by removing the constraint given by the delta function on the free 
energy, that is by setting $u=0$ in the saddle point equations and in the 
complexity. The complexity $\Sigma_{tot}$ that BM find in this way is nontrivial 
(Fig.1 of \cite{bm}). If we set $u=0$ in our equation (\ref{apple}), however, 
we obtain a trivial result, that is $\Sigma=0$. This is consistent with our
initial remarks on the modulus of the determinant: without the constraint
on the free energy we are counting {\it all} the stationary points of the
TAP free energy, each one weighted with the sign of the determinant. This
quantity must be a topological invariant due to the Morse theorem, and thus 
the complexity must vanish. The fact that this happens in our calculation 
and not in the calculation of BM shows that the BRST symmetry selects the 
saddle point solution which preserves Morse's invariance. 

In order to obtain $\Sigma_{tot}$ in our case we simply note that at each 
temperature the dominant value of the complexity is given by its maximal value, 
and therefore $\Sigma_{tot}(T)= \Sigma_{th}(T)$. Thus, the (annealed) average 
global number of solutions of the TAP equations is given by ${\cal N}_{tot}(T)\sim 
\exp(N \Sigma_{th}(T))$. The threshold values $f_{th}$ and $\Sigma_{th}$ can 
be obtained by requiring that the system of equations (\ref{menose}) ceases to 
have solution, i.e. by imposing a marginality condition on their Hessian.
The quantity $\Sigma_{th}(T)$ is plotted in Fig.2: it turns out to be 
considerably smaller than the non-BRST symmetric value of \cite{bm}.
This might be one of the reasons why it is so difficult to find solutions 
of the TAP equations numerically. A numerical analysis of $\Sigma_{tot}(T)$ 
shows that it goes to zero for $\epsilon=T_c-T\to 0$ as,
\beq
\Sigma_{tot}(T) = 0.01\  \epsilon^6 + {\mathrm O}(\epsilon^7) \ , 
\eeq

\begin{figure}
\includegraphics[clip,width=4 in]{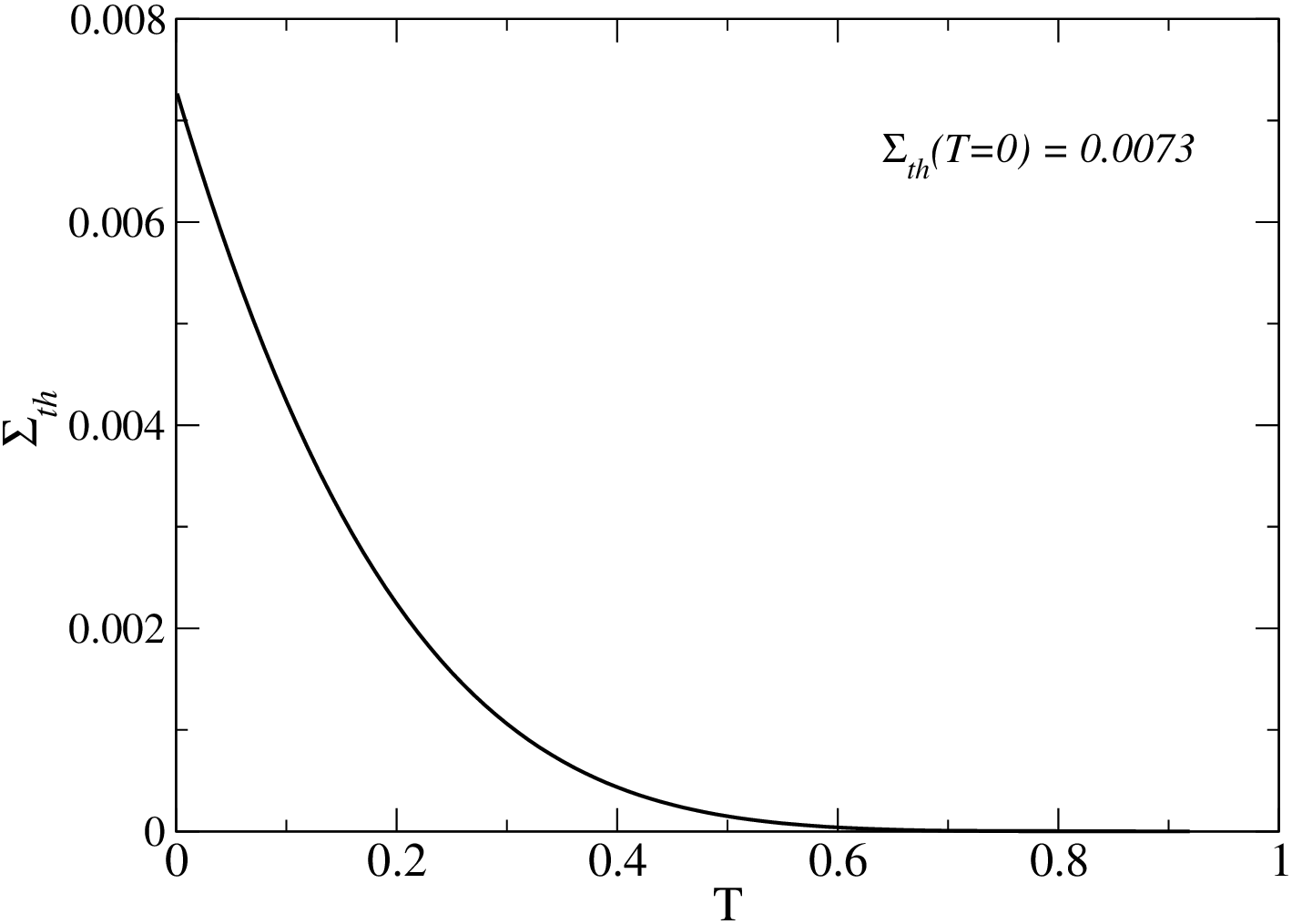}
\caption{The threshold complexity $\Sigma_{th}$ as a function of 
the temperature. Due to the exponential form of the number of states, 
$\Sigma_{th}$ is also the total TAP complexity of the system.}
\label{sigmafig2}
\end{figure}

We conclude this Section with an interesting remark on the comparison 
between the SK and the $p$-spin spherical model. In the $p$-spin model when
we compute the complexity without the constraint on the free energy (i.e.
setting $u=0$) and without the modulus, we find two saddle points: a BRST
symmetric one, which gives $\Sigma=0$, as in the SK case, and a non-BRST 
symmetric saddle point, which gives a nontrivial complexity, $\Sigma'\neq 0$. 
In the $p$-spin spherical model this second value $\Sigma'$ of the complexity 
coincides with $\Sigma_{th}$, that is the maximal value of $\Sigma(f)$ obtained 
by keeping $u\neq 0$ and which is BRST symmetric. The fact that our result, Fig.2, 
does not coincide with the result of BM (Fig.1 of \cite{bm}), shows that this 
identity does not hold in the SK model. Thus, the identity $\Sigma'=\Sigma_{th}$
seems to be a peculiarity of the $p$-spin spherical model.

\subsection{The zero temperature limit of the total complexity}

The largest value of the total complexity $\Sigma_{tot}(T)$ is achieved at $T=0$. 
We find,
\beq
\lim_{T\to 0}\, \Sigma_{tot}(T)= 0.0073 \ .
\label{zotta}
\eeq 
This value is remarkably smaller than the value $\Sigma_{tot}^0=0.199$ 
obtained in \cite{tana,dd1,bm} by directly solving the zero-temperature limit
of the TAP equations. In this Section we want to discuss this difference
and its implications.

The TAP equations at finite $\beta$ are given by,
\beq
m_i= \mathrm{tanh}\{\beta[h_i - \beta(1-q)m_i]\} \ ,
\eeq
where the local field $h_i$ is,
\beq
h_i = \sum_{j\neq i}^N J_{ij} m_j \ .
\eeq
For very large, but finite, $\beta$, we can write,
\beq
m_i = \mathrm{sign}[h_i - \beta(1-q)m_i] \ ,
\eeq
where the magnetizations are now spin variable, $m_i=\pm1$.
In the limit $\beta\to\infty$ we have $q\to 1$, and thus we must be careful with
the expression in the square bracket. Let us write in general for $\beta \gg 1$,
\beq
(1-q)= h_0\beta^{-\alpha}  \quad\quad , \quad\quad \alpha > 0 \ , 
\eeq
such that the TAP equations become,
\beq
m_i= \, \mathrm{sign}\left[h_i - \frac{h_0 \, m_i}{\beta^{\alpha-1}}\right] \ .
\label{zerotap}
\eeq
According to the value of $\alpha$ we have different scenarios.
For $\alpha < 1$ there is no finite limit of the TAP equations (\ref{zerotap}). 
This case is therefore uninteresting. For $\alpha=1$, on the other hand,
we have a well defined zero temperature limit of the TAP equations, i.e
\beq
m_i= \, \mathrm{sign}\left[h_i - h_0 \, m_i\right] \ ,
\label{nostra}
\eeq
This is our case: in figure 3 we show that in the 
limit $T\to 0$ the overlap $q$ corresponding to the threshold (i.e. total) 
complexity behaves like, 
\beq
q= 1 - h_0 T \quad\quad , \quad \quad h_0=0.49 \ ,
\label{049}
\eeq
so that $\alpha=1$. Thus, the zero-temperature limit of the BRST complexity,
$\Sigma_{tot}^0=0.0073$, is the complexity of the solutions of equations 
(\ref{nostra}). Note that the $T$-dependence of the overlap may be different
from (\ref{049}) at values of the free energy different from the threshold.

\begin{figure}
\includegraphics[clip,width=4 in]{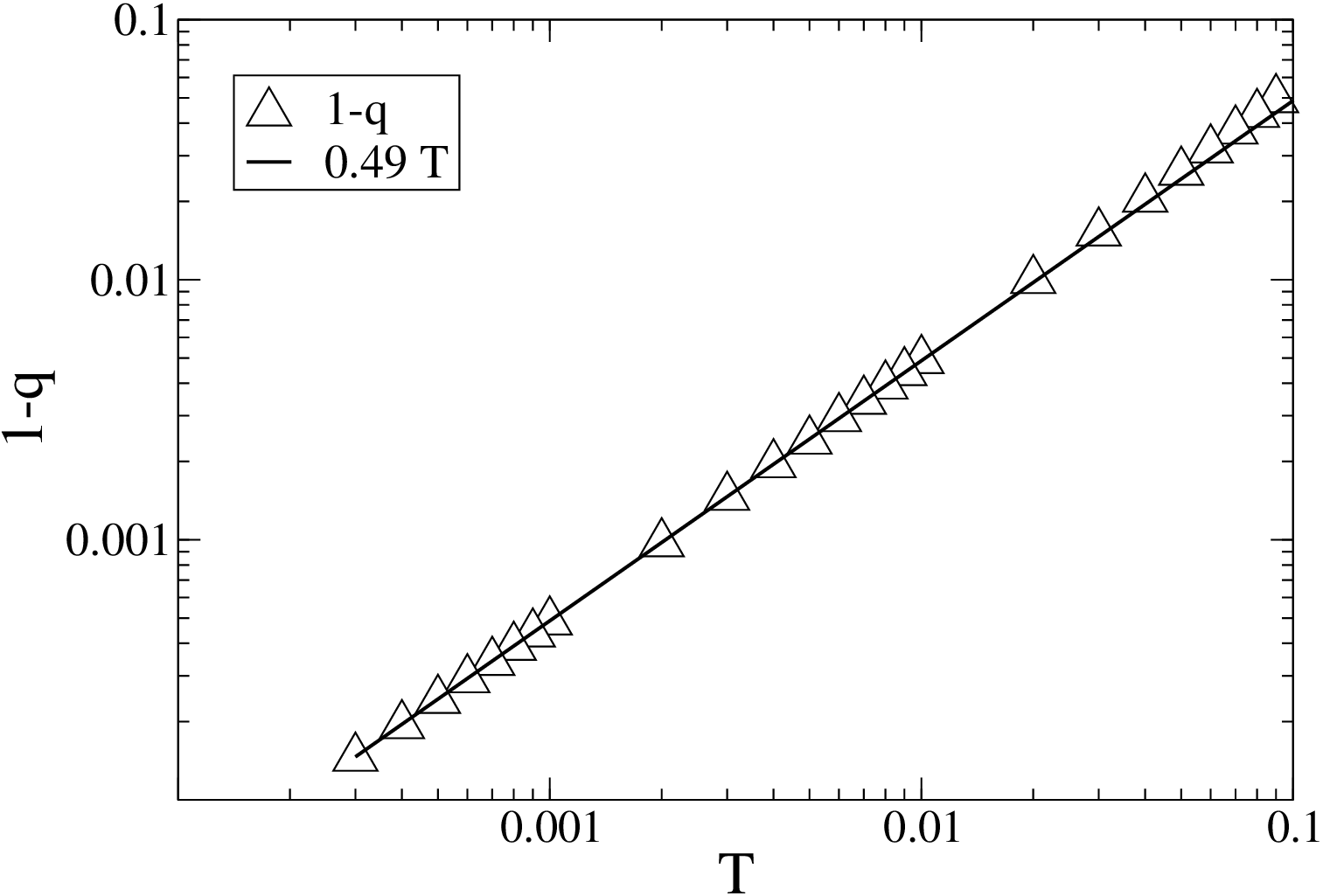}
\caption{The value of the threshold overlap is used to plot $1-q$ as a function of 
$T$. The full line is the best fit to $1-q=h_0 T$, with $h_0=0.49$.}
\label{sigmafig3}
\end{figure}

In the case $\alpha > 1$ the zero temperature limit of the TAP equations 
(\ref{zerotap}) yields,
\beq
m_i= \, \mathrm{sign}\left[h_i\right] \ .
\label{loro}
\eeq
These are the equations considered in \cite{tana, dd1,bm} in order to compute
the zero temperature complexity (in particular, equations (\ref{loro}) were obtained 
by assuming that $q= 1 - O(T^2)$, i.e. $\alpha=2$). 
Therefore, the value $\Sigma_{tot}^0=0.0073$ we obtain as zero-$T$ limit of
the BRST complexity and the value $\Sigma_{tot}^0=0.199$ obtained in 
\cite{tana,dd1,bm}, refer to two different sets of equations, that is 
(\ref{nostra}) and (\ref{loro}). 
The fact that solutions of (\ref{nostra}) have a much smaller complexity 
than those of (\ref{loro}), is due to the fact that the former are actually a 
subset of the second. Indeed, from equations (\ref{nostra}) we have,
\beq
|h_i| > h_0     \ ,
\eeq
for all $i$, which is a very restrictive condition not necessarily satisfied by the
solutions of (\ref{loro}).

We note that equations (\ref{zerotap}) imply,
\beq
m_i h_i > \frac{h_0}{\beta^{\alpha-1}} \ ,
\label{maggio}
\eeq
and thus, given a TAP solution, the change in energy $\Delta E(K)$ when we 
flip $K$ spins, satisfies the relation,
\beq
\Delta E(K) = \sum_i^K m_i h_i -\sum_{ij}^K J_{ij} m_i m_j \sim
\sum_i^K m_i h_i - \frac{K}{\sqrt{N}}\  >  \
K \left( \frac{h_0}{\beta^{\alpha-1}}-\frac{1}{\sqrt{N}}\right) \ .
\eeq
The condition of stability under $K$ spin flips \cite{spinflip} requires $\Delta E(K) > 0$. 
A sufficient (but not necessary) condition for this to hold, is given by,
\beq 
\frac{h_0}{\beta^{\alpha-1}}-\frac{1}{\sqrt{N}} > 0 \ .
\label{stab}
\eeq
In the case $\alpha=1$ this condition is trivially satisfied, while
for $\alpha > 1$ the order of the two limits $\beta\to\infty$ and 
$N\to\infty$ becomes important, and stability under $K$ spin flips 
may be trickier to prove. Nevertheless, a word of caution on the concept of 
stability is in order at this point. 
At finite temperature a stable solution of the TAP equations is 
a local minimum of the TAP free energy, that is a solution with  positive 
Hessian. Unstable saddle points of the TAP free energy will 
exist, typically at high free energies. Unfortunately, at $T=0$ this
topological stability is impossible to assess, because of the discrete 
nature of the model. On the other hand, the $K$-spin flips stability 
is easily defined at $T=0$. However, it is not straightforward to understand what is 
the relation  between these two definitions of stability. We simply 
note that we have computed the total complexity by finding the {\it threshold states}:
for free energies above the threshold value the saddle points equations become unstable. 
Results in continuous mean-field spin-glasses (as the $p$-spin spherical model) 
suggest that the threshold may indeed be the border between topologically stable solutions, 
that is TAP minima, and unstable saddle points. It is therefore plausible
that our total complexity refers only to TAP minima. The connections we 
have discovered between complexity and thermodynamics support this
hypothesis.

\section{Connection with the Legendre transform approach}

As discussed in the previous Section there is a precise correspondence 
between the TAP computation at the lower band edge $f_0$ and the  
computation of the thermodynamic free energy of the system. 
However, the relation between the TAP approach and the usual static one (which
involves Boltzmann averages) is still deeper: as it can be appreciated in 
(\ref{end}) there is a formal connection between the TAP 
complexity at {\it generic} $f$ and the free energy function $F_{1RSB}$.
These equations also represent  an important bridge between the TAP approach  
and some different methods to compute the complexity
which do not rely on TAP equations, but rather on constrained thermodynamics
\cite{monasson,franzparisi}.
In  \cite{monasson}  it is argued that the complexity 
$\Sigma(\beta,f)$ of the ergodic components - i.e. states - present at low 
temperature  is given by the  Legendre  Transform of the free energy 
$ F(\beta,n)$ of $n$ coupled real replicas of 
the original system. More precisely,
\beq
\Sigma(\beta,f)=\max_{n}\ [\beta n\, f - \beta F(\beta,n)] \ ,
\label{monasson}
\eeq
where $n$ and $f$ are Legendre conjugated variables, i.e.
\beq
f=\frac{\partial  F(\beta,n)}{\partial n} \ .
\label{monasson2}
\eeq 
If we assume, as it is generally accepted, that TAP minima correspond
to low temperature states, then the complexity computed via  
eq.(\ref{monasson}) and the complexity obtained from the TAP 
approach should be equal. In other words, we expect that the  
TAP complexity to satisfy eqs. (\ref{monasson}) and (\ref{monasson2}).

To establish this point one has first to compute the free energy 
$ F(\beta,n)$ of the coupled system 
with the  replica method and choose an appropriate Ansatz.  
This problem is discussed for the SK model in \cite{potters}, where 
different kinds of Ansatz are analyzed: within a generalized 
two-group Ansatz for $F(\beta,n)$, equation (\ref{monasson}) 
is formally satisfied by the BM solution, as already noted by these authors in 
\cite{bm,bm1}. Moreover, a simpler sub-class of 
two-group saddle point solutions exists, which corresponds to the following 
procedure:
$F(\beta,n)$ at the $k$-level of RSB is given by $n$ times the 
free energy of one system evaluated with $k+1$ steps of RSB, and breaking 
parameters $n x_1,nx_2,\dots nx_k,n$. Thus, at the lowest possible level of 
RSB ($k=0$) one has,
\beq
F(\beta,n)=n\, F_{1RSB}(\beta;q_1,q_0,n) \ ,
\label{potters}
\eeq
as also suggested in \cite{monasson}. With this expression of 
$F(\beta,n)$, eq. (\ref{monasson}) has been verified within the TAP approach 
for models which are exactly solved by 1RSB replica solution  \cite{potters2,juanpe}. 
It has also been derived for the Bethe Lattice 
at zero temperature with the cavity method \cite{cavity}, and seems 
therefore quite robust.
Surprisingly, for the SK model this point was still unclear, since the BM
solution \cite{bm}, despite its formal correspondence to a more general two-group 
static solution, does not appear to satisfy relation (\ref{monasson}) with the simpler 
Ansatz (\ref{potters}).  The reason for this can be understood in the light of the 
previous Section: in \cite{bm} BM do not use the BRST supersymmetry 
and consequently consider a larger set of solutions than those 
physically relevant, much as the generalized two-group Ansatz has a 
larger set of saddle points than the simpler Ansatz (\ref{potters}). 
We therefore may  expect that 
once the BRST relations are taken into account, as we have done in the previous Section, 
consistency must be recovered.  Indeed, this is precisely what happens: as it can be easily 
verified, once  (\ref{potters}) is implemented, eq.(\ref{end}) is equal to (\ref{monasson}), 
with $q_0=0$, $q_1=q$ and $n=-u$. The parameters $q_1$ and $n$ are variationally fixed:
(\ref{parz2}) is equal to (\ref{monasson2}), while variation with respect to $q_1$ 
gives back equation (\ref{parz1}).

Thus, also for the SK model, we have demonstrated within the TAP approach the validity of
relation  (\ref{monasson}). We note that a crucial element to establish this point is the 
exploitation of the BRST relations. This is not a surprise since, as  discussed in Section III for 
equilibrium averages, the BRST supersymmetry is what mathematically guarantees relations of 
physical significance as (\ref{campo}) and (\ref{energia}).

\section{Conclusions}

In this paper we have discussed the equivalence between the TAP approach and the standard
thermodynamic method for mean-field spin-glass systems, and we have shown that the use of the 
BRST symmetry of the TAP action is crucial to establish this equivalence.

We demonstrated in Section III that the BRST relations guarantee the mathematical 
consistency of the TAP approach and ensure the validity of physically relevant relations. 
We have shown that the DDY computation, which establishes for the SK model a formal 
connection between the TAP and the thermodynamic partition functions, is generally 
valid if the BRST symmetry is taken into consideration, and does not rely on any 
hypothesis about the nature of the equilibrium states.
Furthermore, we have revisited the BM calculation of the complexity for the SK model, with the
support of the BRST relations. Thanks to them, we have been able to solve the saddle point equations
at given free energy, we have explicitly shown the equivalence of the lowest TAP solutions
with the thermodynamic equilibrium states and we have verified the consistency of the 
TAP computation of the complexity with the Legendre Transform construction.

It should be noted that the BRST relations have a very delicate role due to the presence 
of quenched disorder. Indeed, once the average over the disorder is taken and the 
auxiliary variables are introduced, the BRST symmetry of the effective action is much less 
transparent. In the thermodynamic limit, the auxiliary variables become variational parameters which 
must satisfy the saddle point equations. The solutions of these equations 
correspond to averaged quantities (for example, in the BM calculation $R=\langle x_i m_i\rangle$) 
and should therefore satisfy the appropriate BRST relations. However, these relations are 
{\it not} automatically satisfied by all the saddle point solutions. In other words, the set 
of solutions of the saddle point equations in general includes also solutions which are not 
BRST invariant. Thus, the BRST relations must be explicitly imposed, providing some additional 
equations that must be satisfied by the variational parameters. Besides, as we have seen, this 
procedure provides a great simplification in the computation, since it drastically decreases 
the number of unknown variables.

Our BRST calculation gives a total complexity at $T=0$ which is much smaller than the 
value previously obtained by other classic calculations. We have shown that technically
this is due to the fact that the BRST saddle point provides a set of TAP equations at 
$T=0$ which are different from the ones previously considered. In particular, the zero-temperature
BRST-TAP solutions have local fields larger than a finite value $h_0$ at all sites, and 
this condition is responsible for the drastic decrease in the complexity. We have outlined
what are the possible physical explanations of this difference, but more study on the
$T\to 0$ limit is necessary for a complete clarification of this point.

\begin{acknowledgments}
We thank T. S. Grigera, A. Pagnani and F. Ricci-Tersenghi for numerous
interesting discussions, and A. Bray and M. A. Moore for many important 
remarks on the manuscript.
\end{acknowledgments}


\begin{thebibliography}{22}

\bibitem{ea} Edwards S F and Anderson P W 1975 {\it J. Phys. F: Metal. Phys.} {\bf 5} 965.
\bibitem{sk} Sherrington D and Kirkpatrick S 1975 {\it Phys. Rev. Lett.} {\bf 32} 1792.
\bibitem{sk2} Kirkpatrick S and Sherrington D 1978 {\it Phys. Rev. B} {\bf 17} 4384;
\bibitem{1rsb} Parisi G 1979 {\it Phys. Lett.} {\bf 73A} 203.
\bibitem{2rsb} Parisi G 1980 {\it J. Phys. A: Math. Gen.} {\bf 13} L115 and 1101.
\bibitem{tap} Thouless D J, Anderson P W and Palmer R G 1977 {\it Phil. Mag.} {\bf 35} 593.
\bibitem{bm} Bray A J and Moore M A 1980 {\it J. Phys. C: Solid. St. Phys} {\bf 13} L469.
\bibitem{bm1} Bray A J and Moore M A 1980 {\it J. Phys. C: Solid. St. Phys} {\bf 13} L907.
\bibitem{bm2} Bray A J and Moore M A 1980 {\it J. Phys. A: Math. Gen.} {\bf 14} L377.
\bibitem{dd1} De Dominicis C, Gabay M, Garel T and Orland H 1980 {\it J. Physique} {\bf 41} 923.
\bibitem{innocent2} S.A. Roberts 1981 {\it J. Phys. C} {\bf 14}, 3015 (1981).
\bibitem{ddy} De Dominicis C and Young A P, 1983 {\it J. Phys. A: Math. Gen.} {\bf 16} 2063.
\bibitem{MPV} M\'ezard M,  Parisi G and Virasoro M.A., {\it Spin glass theory and beyond'', World Scientific} (1987).
\bibitem{cavity} M\'ezard M and Parisi G 2002 {\it Preprint cond-mat/0207121}.
\bibitem{brs} Becchi C, Rouet A and Stora R 1974 {\it Phys. Rev. Lett.} {\bf 52B} 344.
\bibitem{tito} Tyutin I V 1975 {\it Lebedev preprint} FIAN 39, unpublished. 
\bibitem{crisa1} Crisanti A and Sommers H-J 1992 {\it Z. Phys.} B {\bf 87} 341
\bibitem{kpz} Kurchan J, Parisi G and Virasoro M A 1993 {\it J. Phys. I France} {\bf 3} 1819
\bibitem{crisatap} Crisanti A and Sommers H-J 1995 {\it J. Phys. I France} {\bf 5} 805
\bibitem{monasson} Monasson R 1995 {\it Phys. Rev. Lett.} {\bf 75} 2847.
\bibitem{franzparisi} Franz S. and Parisi G. 1995 {\it J. Physique I} {\bf 3} 1819.
\bibitem{juanpe} Cavagna A, Garrahan J P and Giardina I 1998 {\it J. Phys. A: Math. Gen.} {\bf 32} 711.
\bibitem{potters} Potters M and Parisi G 1995 {\it Europhys. Lett.} {\bf 32} 13.
\bibitem{ps1} Parisi G and Sourlas N 1979 {\it Phys. Rev. Lett.} {\bf 43}, 744.
\bibitem{ps2} Parisi G and Sourlas N 1982 {\it Nucl. Phys. B} {\bf 206}, 321.
\bibitem{zinnjustin} Zinn-Justin J, 1989 {\em Quantum Field Theory and Critical Phenomena}, (Clarendon Press, Oxford).
\bibitem{jorge1} Kurchan J 1991 {\it J. Phys. A: Math. Gen.} {\bf 24} 4969.
\bibitem{jorge2} Kurchan J 2002 {\it Preprint cond-mat/0209399}.
\bibitem{discuss} For an illuminating discussion of the problem of removing the 
determinant \\ in the supersymmetric formalism see \cite{jorge1}. This same problem 
was already \\ taken into consideration in \cite{ps1,ps2}.
\bibitem{mod} Cavagna A, Giardina I and Parisi G  1998 {\it Phys. Rev. B} {\bf 57} 11251.
\bibitem{tana} Tanaka F and Edwards S F 1980 {\it J. Phys. F; Metal Phys.} {\bf 10} 2769.
\bibitem{potters2} Potters M and Parisi G 1995 {\it J. Phys. A: Math and Gen.} {\bf 28} 5267.
\bibitem{spinflip} Biroli G and Monasson R 2000 {\it Europhys. Lett.} {\bf 50} 155.


\end{thebibliography}
\end{document}